# TWO-LAYER QUANTUM KEY DISTRIBUTION

PAULO VINÍCIUS PEREIRA PINHEIRO and RUBENS VIANA RAMOS

paulovpp@gmail.com   rubens.viana@pq.cnpq.br

*Laboratory of Quantum Information Technology, Department of Teleinformatic Engineering, Federal University of Ceara*
*Campus do Pici, C. P. 6007, 60455-740, Fortaleza, Brazil.*

Recently a new quantum key distribution protocol using coherent and thermal states was proposed. In this work, this kind of two-layer QKD protocol is formalized and its security against the most common attacks, including external control and Trojan horse attacks, is discussed.

## 1. Introduction

Quantum key distribution (QKD) has been extensively studied and several protocols were proposed: BB84 [1], Ekert-91 [2], B92 [3], SARG04 [4], continuous variable QKD [5], one-way QKD [6], decoy-state QKD [7], DPS-QKD [8] and MDI-QKD [9] are some examples. The security analysis of such protocols is not trivial. For example, a complete proof of security for DPS-QKD against any type of attack does not exist yet. Furthermore, it seems that all QKD protocols are perfectly secure however, their experimental implementations are not. This happens because, among other problems, the real optical and optoelectronics devices do not follow perfectly their theoretical models used in the security analysis of QKD protocols. Recently, a new step toward secure QKD with real devices, named CT-DPS-QKD, was proposed [10]. Since it is a DPS-QKD protected by a two-state protocol (like B92), it can be seen as a two-layer QKD. The first layer is the QKD protocol while the second layer is a two-state protocol whose goal is exclusively to protect the first layer. The CT-DPS-QKD has some interesting properties. It is secure against beam splitter, intercept-resend, photon number splitting, unambiguous discriminations, external control and Trojan horse attacks. Furthermore, the eavesdropper's presence is denounced in two different ways: parameter (QBER) estimation in the QKD protocol and an electrical signal in the two-state protocol. Additionally, the two protocols are linked in such way that, depending on the attack strategy used by the eavesdropper, attacking one of them causes errors in both. Another interesting point is the fact that the eavesdropper will never be sure about the value of the bits obtained during an attack. Hence, the CT-DPS-QKD is a very robust QKD protocol. The goal of the present work is to formalize the notion of two-layer QKD and to discuss its security, as well to show that different QKD protocols can have their security improved by the use of the second layer.

This work is outlined as follows: In Section 2, we discuss the basic concepts of two-layer QKD. In Section 3 we analyze the possible eavesdropper's strategies against the two-layer QKD. In Section 4 we discuss the distinguishability of quantum light states using a single-photon detector and a spectral analyzer. In Section 5 we describe the two-layer QKD employing differential-quadrature-phase-shift and in Section 6 we discuss its security. In Section 7 we describe the two-layer QKD employing homodyne detection. At last, the conclusions are presented in Section 8.

## 2. Two-layer QKD

Initially, let us remind the prepare-and-measure two-way QKD. Its scheme is shown in Fig. 1.

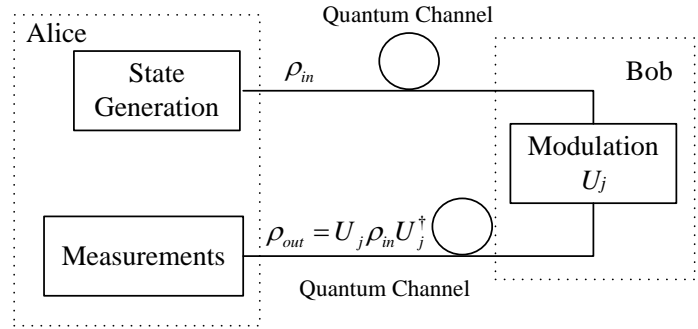

Fig. 1 – Block diagram of the prepare-and-measure two-way quantum key distribution.

The scheme in Fig. 1 works as follows: Firslty, Alice generates and sends to Bob, through a quantum channel, the quantum state $\rho_{in}$, a single-photon or a weak coherent state, for example. Bob, by its turn, using a (polarization or phase) modulator, represented by the unitary operation $U_j$, writes the information $j$ in the quantum state sent by Alice and returns the quantum state $\rho_{out} = U_j \rho_{in} U_j^\dagger$ to Alice. Finally, Alice makes a measurement in the quantum state sent by Bob. The key is obtained from the relation between the modulations applied by Bob (the $j$ values) and Alice's measurements results, as specified by the QKD protocol used (for example, BB84, DPS or B92).

Now, let us consider the following two-state protocol: the sender has two states, $\rho_1$ and $\rho_2$. It sends them to the receiver by using two different optical modes (two time intervals, spatial modes or polarisation modes, for example) and asks the receiver to guess which state is in each mode. Hence, the receiver task is to discriminate between $\rho_1$ and $\rho_2$, having only a single sample of each state. In order to distinguish between the two states with error rate equal to zero, the best strategy for the receiver is to look for a measurement scheme able to distinguish $\rho_1$ and $\rho_2$ unambiguously. However, such scheme does not exist if at least one of the states has a null kernel [10], that is, $\rho_2 \vec{x} = \vec{0} \Rightarrow \vec{x} = \vec{0}$. In this case, the discrimination without errors is impossible. On the other hand, since $\rho_1$ and $\rho_2$ are different quantum states, the statistics of the results of a large number of measurements will be different and, hence, it is possible to distinguish a large sample of $\rho_1$ states from a large sample of $\rho_2$ states.

The two-layer QKD is a two-state quantum protocol (like the one just described) implemented over a prepare-and-measure two-way QKD, as shown in Fig. 2.

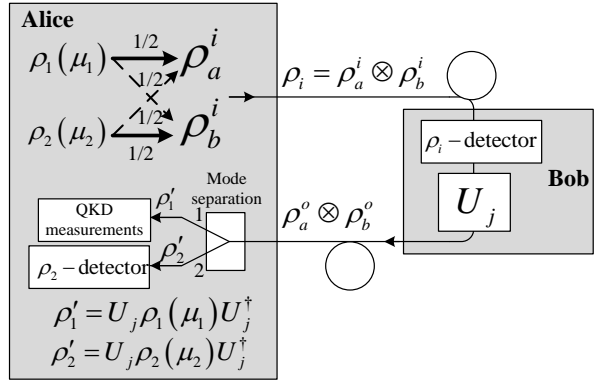

Fig. 2 – Scheme for two-layer quantum key distribution.

In the two-layer QKD, Alice prepares two states $\rho_1$ and $\rho_2$ and sends them to Bob. We will consider that $\rho_1$ and $\rho_2$ are the quantum states used in the two-state protocol, but only $\rho_1$ is used in the QKD protocol. Bob, by its turn, modulates both of them with the same information $j$ and sends back to Alice the quantum states $\rho'_1 = U_j \rho_1 U_j^\dagger$ and $\rho'_2 = U_j \rho_2 U_j^\dagger$. Bob has also a quantum state detector used to prevent a trojan horse attack. A similar procedure is used in QKD schemes using Sagnac interferometer [11]. Once the states $\rho'_1$ and $\rho'_2$ arrive at Alice's place, they are separated: $\rho'_1$ is sent to a measurement setup and the results will be used to form the key; $\rho'_2$ is sent to a another measurement setup used to distinguish between $\rho'_2$ and a quantum state different of $\rho'_2$. If other state not compatible with $\rho'_2$ is measured, the scheme sends an alert signal that denounces the presence of a spy.

In a more detailed way, the scheme showed in Fig. 2 works as follows: Alice has two different quantum states $\rho_1$ and $\rho_2$ (two different quantum light states, like coherent and thermal states). They depend on a parameter $\mu$ (for example, the mean photon number) whose values $\mu_1$ for $\rho_1$ and $\mu_2$ for $\rho_2$ are known only by Alice. She also has two optical modes $a$ and $b$ (like polarization or temporal modes). With probability 1/2 Alice sends $\rho_{1(2)}$ via mode $a$ and $\rho_{2(1)}$ via mode $b$, hence, the total state generated by Alice is $1/2(\rho_1 \otimes \rho_2)_{ab} + 1/2(\rho_2 \otimes \rho_1)_{ab}$. These two states are sent to Bob through a quantum channel (an optical fiber) that, by its turn, applies randomly one of the unitary operations $U_j$ ($j = 0,1$) in both modes and sends them back to Alice. Bob also analyses if the quantum states passing by his devices are in fact the quantum states produced by Alice. Thus, the $\rho_i$-detector consists in a scheme for detection of $\rho_i = \rho_{1(2)} \otimes \rho_{2(1)}$. This prevents a Trojan horse attack in which the eavesdropper Eve sends to Bob different quantum states from those produce by Alice. The total state at Bob's output is $\rho_a^o \otimes \rho_b^o = U_j \rho_a^i U_j^\dagger \otimes U_j \rho_b^i U_j^\dagger$. Since Alice knows which state is in each optical mode, she can always separate them in such way that the quantum states at Alice's output 1 and 2 are always, respectively, $U_j \rho_1(\mu'_1) U_j^\dagger$ and $U_j \rho_2(\mu'_2) U_j^\dagger$. Note that $\mu'_1 \neq \mu_1$ and $\mu'_2 \neq \mu_2$ due to losses in the channel and the $\rho_i$-detection. However, as it will be shown latter, this will not be a problem to Alice because knowing $\mu_{1(2)}$ she can determine $\mu'_{1(2)}$.

Now, it is worth to note that the usage of the quantum state $\rho_2$ is advantageous only if:

1. Eve cannot distinguish perfectly between $\rho_1$ and $\rho_2$.
2. Alice can distinguish between a large sample of $\rho_1$ states and a large sample of $\rho_2$ states.
3. $\rho_2$ does not carry any useful information written by Bob.

Since $\rho_1 \neq \rho_2$, the items 1 and 2 can be satisfied if $\rho_1$ and $\rho_2$ are not orthogonal states. On the other hand, the item 3 is satisfied if $U_j \rho_2 U_j^\dagger = \rho_2$. Now, let $D(\rho_1,\rho_2)$ be a distance measure between the quantum states $\rho_1$ and $\rho_2$. If $D(\rho_1,\rho_2) = D_{min}$ (the minimal value of $D$) then $\rho_1$ and $\rho_2$ are indistinguishable by any measurement process. On the other hand, if $D(\rho_1,\rho_2) = D_{max}$ (the maximal value of $D$) then there exist a measurement process able to distinguish perfectly $\rho_1$ and $\rho_2$ with a single measurement. At last, if $D_{min} < D(\rho_1,\rho_2) < D_{max}$ it is possible to distinguish $\rho_1$ and $\rho_2$ if one has a large enough number of samples of one of them. The lower the value of $D$ the larger is the number of samples required. Considering the distance measure $D$ and the scheme in Fig. 2, the requirements that must be obeyed by $\rho_1$ and $\rho_2$ in order to have secure QKD are

$$D(\rho_1, \rho_2) = \varepsilon, \quad D_{min} < \varepsilon \ll D_{max} \tag{1}$$

$$D(U_j \rho_1 U_j^\dagger, U_j \rho_2 U_j^\dagger) = \varepsilon, \; j = 0,1 \tag{2}$$

$$D(\rho_2, U_j \rho_2 U_j^\dagger) = D(\rho_2, \rho_2) = D_{min}, \; j = 0,1 \tag{3}$$

$$\rho_2 \vec{x} = \vec{0} \Rightarrow \vec{x} = \vec{0}. \tag{4}$$

Equation (1) states that it is possible to distinguish a large amount of sample of $\rho_1$ from a large amount of samples of $\rho_2$. Since Eve is forced to make individual attacks (she does not know which state is in each mode), she has to try to distinguish between $\rho_1$ or $\rho_2$ using only one sample of them. In this case, the best strategy for Eve would be to apply an unambiguous discrimination, however, as stated by (4), $\rho_2$ have a null kernel, hence, this attack is not possible. Since the discrimination between $\rho_1$ and $\rho_2$ is not perfect, Eve will never be sure if she is attacking $\rho_1$ or $\rho_2$. This point is crucial for the improvement of the security of the QKD protocol protected by the two-state protocol. Equation (2) states that, after Bob's action, the distinguishability between the states does not change: $D(\rho_a^o, \rho_b^o) = D(\rho_a^i, \rho_b^i)$. Equation (3) states that Bob's action on $\rho_2$ is not noted by anyone, that is, only $\rho_1$ carries the useful information.

Since Alice is always able to separate the modes correctly, the state $\rho_2$ will never appear at output 1 and, hence, it will not cause errors in the QKD protocol. Simultaneously, the state $\rho_1$ will never appear at output 2 and, hence, it will not cause errors in the $\rho_2$ detection scheme.

## 3. The eavesdropper's strategies

The Eve's goal is to determine which $U_j$ was used by Bob without causing any error in Alice and Bob. Here we are going to discuss four kinds of attack named type I, II, III and IV. In the type I attack, Eve will attack the states leaving Bob. Without consider any particular strategy of attack, after Eve's action, the quantum states arriving at Alice's place are

$$\rho_3^j \otimes \rho_4^j = E\left(U_j \rho_a^i U_j^\dagger\right) \otimes E\left(U_j \rho_b^i U_j^\dagger\right). \tag{5}$$

In (5) $E$ is the operator that models Eve's attack. During Alice's mode separation (in average) half of the time $\rho_3^j$ will appear at output 1 and $\rho_4^j$ at output 2 and vice-versa. Therefore, the quantum states at Alice's output 1 and 2 are

$$\left[\frac{1}{2}\rho_3^j + \frac{1}{2}\rho_4^j\right]_1 \otimes \left[\frac{1}{2}\rho_3^j + \frac{1}{2}\rho_4^j\right]_2. \tag{6}$$

In order to avoid errors in Alice, the following conditions on $\rho_3^j$ and $\rho_4^j$ must be satisfied

$$D\left(\frac{1}{2}\rho_3^j + \frac{1}{2}\rho_4^j, U_j \rho_1 U_j^\dagger\right) = D_{min} \tag{7}$$

$$D\left(\frac{1}{2}\rho_3^j + \frac{1}{2}\rho_4^j, U_j \rho_2 U_j^\dagger\right) = D\left(\frac{1}{2}\rho_3^j + \frac{1}{2}\rho_4^j, \rho_2\right) = D_{min} \tag{8}$$

In other words, the states in (6) must be indistinguishable of the states sent by Bob. If condition (7) is not obeyed there will be errors in the QKD protocol. If condition (8) is not obeyed there will be errors in the two-state protocol. However, the conditions (7) and (8) cannot be satisfied simultaneously because they imply $D\left(U_j \rho_1 U_j^\dagger, U_j \rho_2 U_j^\dagger\right) = D_{min}$ that is not in accordance with (2). In this case there will be errors in both protocols.

In the type II attack, Eve tries to identify which state is in each mode before Bob's actions, that is, Eve tries to distinguish between $\rho_1$ and $\rho_2$. Once $\rho_1$ was identified, attacks of the type I are applied exclusively on it. However, due to (1) Eve cannot do this, taking each state individually, without errors or without a degree of uncertainty. As explained before, the best strategy to Eve would be to apply an unambiguous discrimination, however, due to (4), $\rho_1$ and $\rho_2$ cannot be discriminated unambiguously.

In the type III attack, Eve applies a Trojan horse attack. She stores in a quantum memory the states sent by Alice, prepares her own states, in general quantum states different from those used by Alice, and sends them to Bob. At Bob's output, Eve recovers the states and measure them aiming to obtain some information about the quantum operation realized by Bob. After, according to the results of her measurements, Eve applies a unitary operation in both states sent by Alice and sends them back to Alice. In order to prevent this attack, as can be seen in Fig. 2, Bob uses a detector that informs if the states processed by him are compatible with the states sent by Alice. The crucial point here is the parameter $\mu$ whose values are known only by Alice. At the end of the quantum communication Alice informs to Bob the values of $\mu$ used for $\rho_1$ and $\rho_2$, and Bob checks if his measurement results are in accordance with those $\mu$ values. Since Eve does not know $\mu_1$ and $\mu_2$ she will have to guess which quantum state to use in order to realize the attack without causing an alert signal in Bob. Since $\mu$ is a continuous variable, this task will be very hard to Eve. Without changing the quantum state processed by Bob, the type III attack is not useful for Eve.

In the type IV attack, Eve tries to control Alice's measurement setup by sending a quantum state, prepared by her, different from $\rho_1$, let us call it $\rho_E$. However, since Eve does not know in which mode $\rho_1$ is, sometimes the quantum state $\rho_E$ will appear at output 2, hence, the state $\rho_E$ will be measured by the scheme able to identify $\rho_2$, and an error sign will be fired, informing Eve's presence.

Summarizing, attacks type I, II and IV will always cause errors in Alice while type III will cause errors in Bob. Furthermore, Eve will not be sure about the bit values obtained in her attacks because she does not know if she is attacking $\rho_1$ or $\rho_2$ (type I and II attacks), she cannot change the quantum states sent to Bob (type III) and she cannot to control externally the bit value received by Alice (type IV).

## 4. Distinguishability of quantum light states by using a single-photon detector and a spectrum analyzer

The first two-layer QKD protocol was proposed in [12]. It was named CT-DPS-QKD because the QKD protocol used is the DPS-QKD and the quantum states used in the two-state protocol are the coherent state (also used in the QKD protocol) and the thermal state, thus

$$\rho_1 \equiv \rho_\alpha = |\alpha\rangle\langle\alpha|, \quad |\alpha\rangle = \sum_{n=0}^{\infty} \exp\left(-|\alpha|^2/2\right)\frac{\alpha^n}{\sqrt{n!}}|n\rangle \quad (9)$$

$$\rho_2 \equiv \rho_t = \left[\mu_t^n/(1+\mu_t)^{1+n}\right]|n\rangle\langle n|. \quad (10)$$

The parameter $\mu$ is the mean photon number: $\mu_1=|\alpha|^2$ and $\mu_2=\mu_t$. The overlap between $\rho_\alpha$ and $\rho_T$ is given by

$$\langle\alpha|\rho_t|\alpha\rangle = \exp\left[-|\alpha|^2/(1+\mu_t)\right]/(1+\mu_t). \quad (11)$$

Since $\rho_\alpha$ and $\rho_T$ are not orthogonal for finite values of $\mu_t$ and $|\alpha|^2$, they cannot be perfectly distinguished with a single measurement. However, if $\mu_t+|\alpha|^2\neq 0$, the states $\rho_\alpha$ and $\rho_t$ can be distinguished with high probability if one has a large enough number of samples of one of them. As explained in [12,13] this task can be realized by a threshold single-photon detector. Ignoring the afterpulsing, the probabilities of the thermal and coherent states to fire an avalanche in a threshold single-photon detector are, respectively, given by

$$P_t = 1 - \left[1/(1+\eta\mu_t)\right](1-p_d) \quad (12)$$

$$P_c = 1 - \exp(-\eta|\alpha|^2)(1-p_d). \quad (13)$$

In (12) and (13), $\eta$ and $p_d$ are, respectively, the single-photon detector quantum efficiency and dark count probability. Using a spectrum analyser to measure, in a fixed frequency band, the electrical power of the signal produced by a threshold single-photon detector, a large sample of thermal states can be distinguished of a large sample of coherent states. This happens because the electrical power in a fixed band is proportional to $(P-P^2)$ where $P$ is the probability of an avalanche to be fired [13]. Since $|\alpha|^2$ and $\mu_t$ are chosen such that the probabilities in (12) and (13) are different (but still keeping the condition (1) satisfied), the electrical powers measured will also be different. Applying this strategy, Alice and Bob can check if the quantum states they are receiving are in fact those they were expecting to receive. Considering the two-layer QKD protocol here discussed, in general any other quantum state having the vacuum probability different from the vacuum probability of the thermal state can be distinguished from the thermal state if a sufficient larger number of samples are provided.

Considering the usage of the quantum light discriminator using a threshold single-photon detector and a spectrum analyzer in the two-way QKD protocol, once Alice chose the value of $\mu_t$, knowing the losses in the optical link, she can infer the optical power she will measure in a given frequency band. If she measures a value different from that one expected, she aborts the key distribution. Hence, any Eve's attack that changes the electrical power measured by Alice will denounce her presence. That is why type I, II and IV attacks cannot be used by Eve: all of them modify the quantum state that arrives at Alice's output 2, the quantum state that will be measured by the quantum light discriminator composed by a single-photon detector and a spectrum analyzer.

A similar quantum light discriminator is also used by Bob. The goal is to avoid the type III attack. For example, Eve could stop the coherent and thermal states sent by Alice, to keep them in a quantum memory and sending two single-photons to Bob, $|1\rangle|1\rangle$. After Bob's modulation, at Bob's output Eve gets two photons with the same information. However, the usage of the states $|1\rangle|1\rangle$ instead of $\rho_\alpha \otimes \rho_t$ will result in an electrical power (measured by the spectral analyzer) different from that one expected by Bob.

## 5. The CT-DQPS-QKD protocol

In this section we are going to consider the two-layer QKD employing the differential-quadrature-phase-shift QKD [14]. The optical scheme used to implement the CT-DQPS-QKD is shown in Fig. 3.

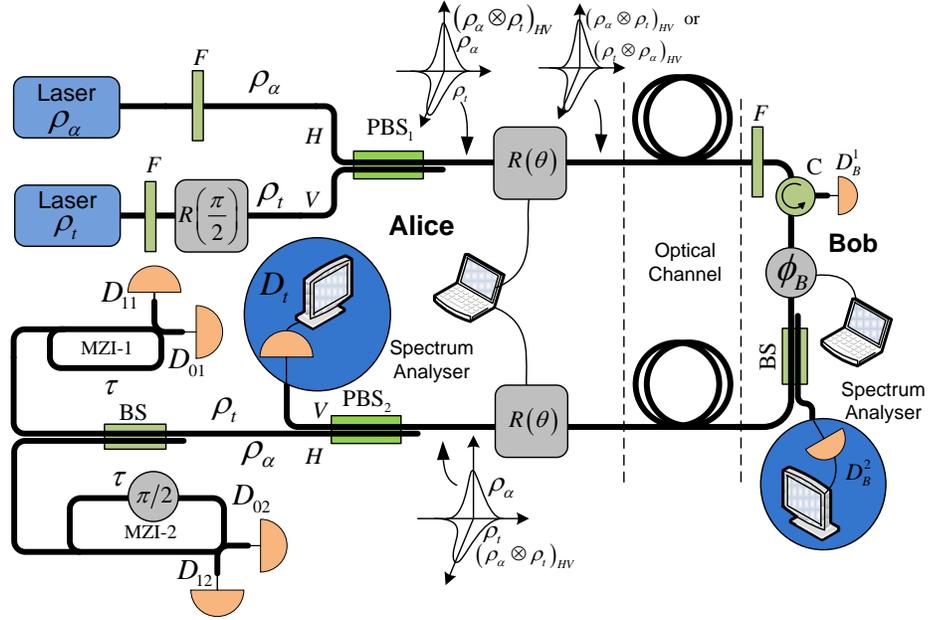

Fig. 3 - Optical setup for implementation of the two-layer CT-DQPS-QKD. PBS – polarizing beam splitter; R – polarization rotator; $\phi_B$ – phase modulator; C- optical circulator; F – optical filter with central frequency at 1550 nm; BS - beam splitter; $D_{11}$, $D_{01}$, $D_{12}$, $D_{02}$, $D_t$, $D_B$ (1 and 2) are single-photon detectors.

Comparing the scheme in Fig. 3 with the description of two-layer QKD in Section 2, one can note that we chose $\rho_1 = |\alpha\rangle\langle\alpha|$ and $\rho_2 = \rho_T$. Futhermore, the unitary operation $U$ is a phase-modulator, hence $U = e^{i\phi\hat{N}}$, where $\hat{N}$ is the Number operator. One can readily observe that

$$U\rho_1 U^\dagger = e^{i\phi\hat{N}}|\alpha\rangle\langle\alpha|e^{-i\phi\hat{N}} = |\alpha e^{i\phi}\rangle\langle\alpha e^{i\phi}| \qquad (14)$$

$$U\rho_2 U^\dagger = e^{i\phi\hat{N}}\rho_T e^{-i\phi\hat{N}} = \rho_T. \qquad (15)$$

Now, using (11) one has that one has that the distinguishability between $|\alpha\rangle\langle\alpha|$ and $\rho_T$ and between $U(\phi)|\alpha\rangle\langle\alpha|U^\dagger(\phi)$ and $U(\phi)\rho_T U^\dagger(\phi)$ are the same

$$\langle\alpha e^{i\phi}|\rho_t|\alpha e^{i\phi}\rangle = \langle\alpha|\rho_t|\alpha\rangle = \exp\left[-|\alpha|^2/(1+\mu_t)\right]/(1+\mu_t). \qquad (16)$$

Hence, the pair coherent and thermal states obeys the conditions (1) and (2). Moreover, due to (15) (3) is also satisfied. Finally, since the density matrix of the thermal state is a diagonal matrix with positive entries, it has a null kernel and, hence, (4) is also satisfied. Furthermore, *a* and *b* modes discussed in Section 2 are the horizontal and vertical polarization modes. The mode separation in Alice is simply realized by a polarizing beam splitter. Finally, the $\rho$-detector is the scheme with single-photon detector and spectrum analyzer discussed in Section 4. Initially, Alice produces optical pulses having a coherent state at the horizontal mode and a thermal state at the vertical mode, both of them having low mean photon number such that (1) is satisfied. Two optical filters (*F*) in Alice put the light from both sources inside the same spectral range. This avoids a side-channel attack in which Eve looks for photons in different frequencies aiming to identify from which source the photon came from. Following, Alice, randomly, sets her polarisation rotator $R(\theta)$ in $\theta = 0$ or $\theta = \pi/2$. Thus, for each optical pulse produced by Alice, the total quantum state entering the optical channel is $1/2(\rho_\alpha \otimes \rho_t)_{HV} + 1/2(\rho_t \otimes \rho_\alpha)_{HV}$. Bob has a polarisation insensitive phase modulator [11]. Bob's phase modulation does not change the thermal state (since it has a diagonal density matrix), as is required by (3), but it adds the phase $\phi_B$ to the coherent state. Leaving Bob's place, the optical pulses return to Alice. For each pulse arriving, Alice applies the same polarisation rotation she had applied when the pulse was leaving her place. Thus, before $PBS_2$, all pulses will have the coherent state at *H*-mode and thermal state at *V*-mode. These modes are separated by $PBS_2$ and the thermal state at the *V*-mode is monitored by a single-photon detector plugged to a spectrum analyser that will measure the electrical power in a fixed band. On the other hand, the coherent state at *H*-mode is sent to Alice's QKD measurement setup composed by two fibre interferometers whose time difference between upper and lower arms, $\tau$, is equal to the time separation between two consecutive pulses. The DQPS-QKD protocol can be readily implemented if Bob and Alice play the opposite roles as happens in traditional DQPS-QKD. Thus, Bob modulates each pulse that arrives at his place applying randomly one of the phases 0, $\pi/2$, $\pi$ and $3\pi/2$. Alice, by its turn, is the one who has the interferometers placed at the coherent state output. The protocol rules are the same and its security is increased by the use of thermal states.

## 6. Security of the CT-DQPS-QKD protocol

The security of the CT-DQPS-QKD can be explained as follows: Without knowing if the coherent state is in the *H*- or *V*-mode, the beam splitter, intercept-resend and photon number splitting attacks will not be useful for Eve, since she will not be sure about the bit valued obtained during the attack and she will introduce errors in Alice [12]. All of these are type I attacks. Since the thermal state has a null kernel, the type II attack with unambiguous discrimination does not work.

As can be seen in Fig. 2, the filter *F* in Bob avoids Eve to use light in a different wavelength not detected by Bob's single-photon detector $D_B^2$. The optical circulator and the single-photon detector $D_B^1$ are used to avoid a bidirectional path. At last, the beam splitter BS with reflectance *r*, the single-photon detector $D_B^2$ and the spectrum analyser are used to check if the states modulated by Bob are in fact compatible the coherent and thermal states with the mean photon numbers chosen by Alice. After the quantum communication, Alice informs to Bob the mean photon numbers used. Bob checks if the electrical power value measured during the quantum communication is in accordance with the expected value for those mean photon numbers announced by Alice. If it is not, the key exchanged is discarded. Thus, if Eve changes the quantum states, without knowing the mean photon numbers used by Alice, with high probability she will change the electrical power measured by Bob. Moreover, since the QKD protocol used is the DQPS-QKD, in order to be sure about the quantum operation used by Bob, Eve has to use an optical pulse with at least two photons (if Eve sends to Bob the quantum state $|1\rangle$ the pulse at Bob's output will be the same produced by an ideal BB84). However, since Alice uses low mean photon numbers for the thermal and coherent states, the probability of having two photons in a pulse produced by her will be low. Thus, a two-photon state $|1\rangle_H|1\rangle_V$ will be easily distinguished from the state $1/2(\rho_a \otimes \rho_t)_{HV} + 1/2(\rho_t \otimes \rho_a)_{HV}$ produced by Alice. This is the type III attack. At last, if Eve tries to control externally Alice's single-photon detectors ($D_{01}$, $D_{11}$, $D_{02}$, $D_{12}$) sending to Alice strong (pulsed or CW) light [15], part of this light will be guided to the output 2 and the electrical power measured by Alice will be different from the expected value, indicating the attack. This is the attack type IV.

A simplified analysis of the secret-key rate ($R_S$) of the two-layer CT-DQPS-QKD consists in to consider (unrealistically) that, after her measurements, Eve knows when she attacked a thermal or a coherent state. In this case, one has $R_S = R_{sift}(I_{AB} - I_{AE}/2)$. The term ½ is due to the thermal states: having to choose randomly which mode to attack (horizontal or vertical), in average, half of the bits got by Eve (those obtained when a thermal state is attacked) will be completely uncorrelated with Alice's sequence, implying in $I_{AE} = 0$. The other half of the bits (those obtained when a coherent state is attacked) will imply in $I_{AE} \neq 0$. Using the equations for $I_{AB}$ and $I_{AE}$ described in [16] one has

$$I_{AB} = 1 - \eta_{ec} H(QBER) \tag{17}$$

$$I_{AE} = \left\{(1-\mu/2t)[1-H(p)]+(\mu/2t)\right\}/\left\{1+[2p_d/(\mu t \eta)]\right\} \tag{18}$$

$$p_\mu = \mu t \eta t_B \tag{19}$$

$$t = 10^{-\alpha L/10} \tag{20}$$

$$R_{sift} = \left[1/2\left(p_\mu + 2p_d + p_{ap}\right)f_{rep}\right]/\left[1+\tau_{dead}f_{rep}\left(p_\mu + 2p_d + p_{ap}\right)\right] \tag{21}$$

$$p_{ap} = 0.008\left(p_\mu + 2p_d\right) \tag{22}$$

$$QBER = \frac{1}{2}\frac{p_\mu(1-V)+2p_d+p_{ap}}{p_\mu+2p_d+p_{ap}} \tag{23}$$

$$p = 1/2 + \sqrt{D(1-D)} \tag{24}$$

$$D = (1-V)/(2-\mu/t). \tag{25}$$

In (17)-(25), $\eta_{ec}$ (=1.2) is the efficiency of the error correction protocol, $\mu$ (=0.1) is the mean photon number of the coherent state used, $p_d$ (=5.10$^{-6}$) and $p_{ap}$ are, respectively, the single-photon detector's dark count and afterpulsing probabilities, $\eta$ (=0.07) is the single-photon detector's quantum efficiency, $t$ is the transmissivity of the optical link of length $L$ between Alice and Bob, $t_b$ (=0.543) takes into account the losses in Bob devices, $\alpha$ (=0.21 dB/km) is the fibre coefficient loss, $f_{rep}$ = (5 MHz) is the pulse repetition frequency, $V$ (=0.98) is the visibility of the interferometers, $\tau_{dead}$ (=1$\mu$s) is the time during which the single-photon detector is unable to fire a new avalanche. At last, $H(\cdot)$ is the entropic function: $H(x)=-x\log_2(x)-(1-x)\log_2(1-x)$. The comparison of the $R_{sec}$ for DQPS-QKD with and without the second layer (thermal states) is shown in Fig. 4.

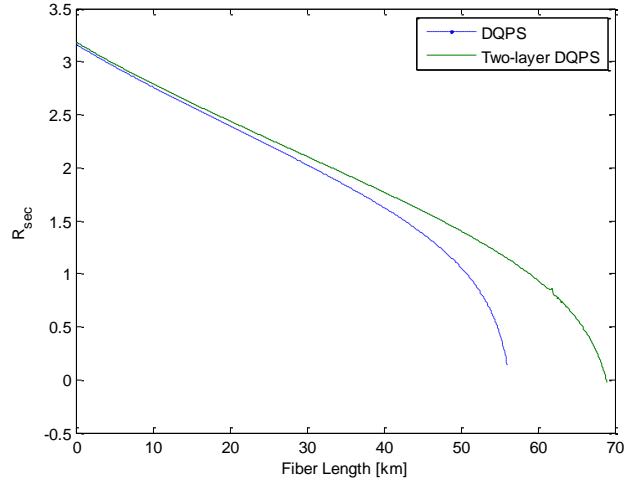

Fig. 4 – $R_{sec}$ [bit/s] (logarithmic scale) versus $L$ for DQPS-QKD and two-layers DQPS-QKD.

In the real case, Eve is never sure if she attacked a thermal or a coherent state. In this situation, the real mutual information between Alice and Eve will be lower than $I_{AE}/2$ and the curve of the secret-key rate will be over those shown in Fig. 4.

## 7. Two-layer QKD using homodyne detection

Although up to now we have discussed only the two-layer QKD employing the DQPS-QKD protocol, the usage of two layers for other QKD protocols is straightforward, for example, see [17] for a two-layer QKD using one-way QKD. Here, we briefly present the scheme for two-layer QKD employing the QKD protocol with homodyne detection presented in [18]. The proposed scheme is shown in Fig. 5.

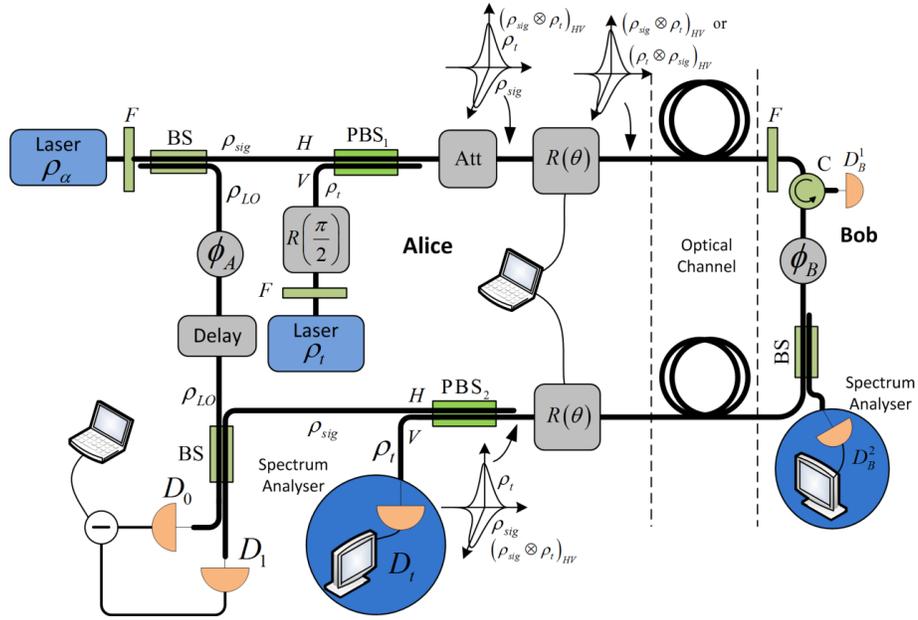

Fig. 5 – Scheme for a two-layer quantum key distribution using homodyne detection.

The protocol can be understood as follows: The light emitted by Alice's coherent source is split by an unbalanced beam splitter into the signal ($\rho_{sig}$ - with mean photon number lower than 1 after the optical attenuator) and local ($\rho_{LO}$ - with mean photon number around $10^6$) components. Alice applies randomly a phase shift $\phi_A$ (0, $\pi$/2) in $\rho_{LO}$. The delay line is used to compensate the time needed to the signal pulse sent to Bob to return to Alice's apparatus. Once in Bob, the light beams are randomly phase shifted (0, $\pi$/2, $\pi$, 3$\pi$/2). Once again, the thermal state is not affected by Bob's phase modulator. At Alice's apparatus input, the same polarization rotation must be applied in order to assure the thermal state in the vertical polarization and the signal (coherent) pulse in the horizontal polarization. The former is sent to the state analyzer (single-photon detector and spectrum analyzer) and the second is sent to a balanced beam splitter where it will suffer interference with the local pulse phase-modulated by Alice. The difference of the photocurrents produced in $D_0$ and $D_1$ is used to determine the bit value. Let $x$ be the value of difference of the photocurrents measured, $X_D$ and $-X_D$ are reference values and $\phi = \phi_B - \phi_A$. If $-X_D < x < X_D$ then the result is inconclusive and no bit is obtained. On the other hand, if $x < -X_D$ or $x > X_D$, the following codification can be used to produce a key: $\phi_A = 0 \rightarrow$ bit '0'; $\phi_A = \pi/2 \rightarrow$ bit '1'; $\phi_B = 0$ or $\pi \rightarrow$ bit '0'; $\phi_B = \pi/2$ or 3$\pi$/2$\rightarrow$ bit '1'. Hence, after having received all the optical pulses, Alice has to inform to Bob the time slots where she obtained a valid measurement result. At last, as expected, the security of the QKD protocol with homodyne detection is improved by the presence of the thermal states.

## 8. Conclusions

A practical quantum key distribution setup can be considered secure if 1) the quantum signals used are of the type required by the protocol's rules; 2) There are not side-channels; 3) Trojan horses and external control of Alice's and Bob's setups are not possible. Considering the two-layer protocol here discussed, it is largely known that coherent and thermal quantum states are well approximated, respectively, by the light produced by a laser operating well above the threshold and by the light produced by a LED or a semiconductor laser operating well below the threshold. Hence, the quantum states required are easily produced by common (non-ideal) optoelectronic devices. The main point here is the time separation between two consecutive pulses sent by Alice. It must be larger than the coherence time of the thermal source. Hence, condition 1) is satisfied and this implies that the protocol's rules, (1)-(4), are also satisfied when real optoelectronic devices are used. If the coherent and thermal optical pulses produced by Alice have the same spectral range, a side-channel attack where Eve tries to determine from which source the light came from using filters is not possible. This is guaranteed by the optical filters used by Alice. Trojan horses attacks and the external control of Alice's and Bob's setups are not possible because of the secrets known only by Alice: the mean photon numbers used and the polarization of the coherent and thermal states. The two main advantages of using the two-layer protocol are: 1) the presence of Eve is detected in two different ways: the QBER estimation in the QKD protocol and by the quantum state analyzer (single-photon detector and spectrum analyzer). 2) Eve is never sure about the bit value she obtained during an attack. Thus, the usage of thermal states decreases the value of $I(A{:}E)$ what results in a improvement of the secret-key rate. Hence, the two-layer protocol is a useful strategy to improve the security of real QKD setups.

## Acknowledgements

This work was supported by the Brazilian agencies CNPq via Grant no. 303514/2008-6. Also, this work was performed as part of the Brazilian National Institute of Science and Technology for Quantum Information.

## References

1. Bennet, C. H.: Brassard, G.: Quantum cryptography: public key distribution and coin tossing, in Proceedings of IEEE International Conference on Computers, Systems and Signal Processing, 175 (1984).
2. Ekert, A. K.: Quantum cryptography based on Bell's theorem, Phys. Rev. Lett., 67, 661 (1991).
3. Bennet, C. H.: Quantum Cryptography Using Any Two Nonorthogonal States, Phys. Rev. Lett., 68, 21, 3121 (1992).
4. Scarani, V., Acín, A., Ribordy, G., and Gisin, N.: Quantum Cryptography Protocols Robust against Photon Number Splitting Attacks for Weak Laser Pulse Implementations, Phys. Rev. Lett., 92, 057901 (2004).
5. Grosshans, F., and Grangier, P.: Continuous Variable Quantum Cryptography Using Coherent State, Phys. Rev. Lett., 88, 057902 (2002).
6. Stucki, D., Brunner, N., Gisin, N., Scarani, V., and Zbinden, H.: Fast and simple one-way quantum key distribution, Appl. Phys. Lett., 87, 194108 (2005).
7. Lo, H.-K., Ma, X., and Chen, K.: Decoy state quantum key distribution, Phys. Rev. Lett., 94, 23, 230504 (2005).

8. Inoue, K., Waks, E., and Yamamoto, Y.: Differential-phase-shift quantum key distribution using coherent light, Phys. Rev. A, 68, 2, 022317 (2003).
9. Lo, H.-K., Curty, M., and Qi, B.: Measurement device independent quantum key distribution, quant-ph arXiv:1109.1473 (2011).
10. Rudolph, T., Spekkens R. W., and Turner, P. S.: Unambiguous discrimination of mixed states, Phys. Rev. A, 68, 010301(R) (2003).
11. B. Qi, L.-L. Huang, H.-K. Lo, and L. Qian (2006), Polarization insensitive phase modulator for quantum cryptosystems, Opt. Express 14, 4264-4269.
12. Mendonça, F. A., Brito, D. B. de, Ramos, R. V.: An Optical Scheme for Quantum Multi-Service Network, Quant. Inf. & Comp. 12, 7-8, 620 (2012).
13. Cavalcanti, M. D. S., Mendonça, F. A., and Ramos, R. V.: Spectral method for characterization of avalanche photodiode working as single-photon detector, Opt. Letts., 36, 17, 3446 (2011).
14. Inoue, K. and Iwai, Y., Differential-quadrature-phase-shift quantum key distribution, Phys. Rev. A, 79, 022319 (2009).
15. Lydersen. L. et al: Hacking commercial quantum cryptography systems by tailored bright illumination, Nature Photonics, 4, 686 (2010).
16. Eraerds, P., Walenta, N., Legré, M., Gisin N., and Zbiden, H.: Quantum key distribution and 1 Gbps data encryption over a single fibre, New. J. of. Phys., 12, 063027 (2010).
17. Pinheiro, P. V. P. and Ramos, R. V.: Two-layer one-way quantum key distribution, IV Workshop-school on Quantum Computation and Information, Fortaleza-Brazil (2012).
18. Hirano, T., Yamanaka, H., Ashikaga, M., Konishi, T. and Namiki, R.: Quantum cryptography using pulsed homodyne detection, Phys. Rev. A, 68, 042331 (2003).